\documentclass[10pt]{iopart}

\usepackage{amsgen}
\usepackage{amsfonts}
\usepackage{amsbsy}
\usepackage{amssymb}

\usepackage{tabularx}
\usepackage{graphicx}
\usepackage{times}
\usepackage{amssymb}
\usepackage{hyperref}
\usepackage{textcomp}
\usepackage{xcolor}
\usepackage[T1]{fontenc}
\usepackage{ulem}

\newcommand{\tno} {\,^\circ_\circ\,}

\begin{document}
\title{Enhanced squeezing by absorption}
\author{P. Gr\"unwald and W. Vogel}
\address{Institut f\"ur Physik, Universit\"at Rostock, D-18055 Rostock, Germany}
\ead{Electronic address: peter.gruenwald@uni-rostock.de}

\date{\today}

\begin{abstract}
Absorption is usually expected to be detrimental to quantum coherence effects. However, the situation for complex absorption spectra has been little studied yet. We consider the resonance fluorescence of excitons in a semiconductor quantum well. The creation of excitons requires absorption of the incoming pump-laser light. Thus, the absorption spectrum of the medium acts as a spectral filter for the emitted light. Surprizingly, absorption can even improve quantum effects, as is demonstrated for the squeezing of the resonance fluorescence of the quantum-well system. This effect can be explained by an improved phase matching due to absorption.
\end{abstract}

\maketitle

\ioptwocol

\section{Introduction}\label{sec.intro}
Light fields passing through a medium are generally subject to dispersion and absorption. These effects are a result of internal resonances of the medium. If the light field has a narrow spectral width, or the medium is far from resonance, transmission yields a scaling of the incoming fields. In such scenarios, correlations of fields are diminished by a constant factor, suppressing the magnitude of quantum effects correspondingly. It is well known that spectrally broad absorption reduces squeezing~\cite{Schnabel}. However, fluctuating losses may even preserve quantum effects much better than standard attenuating, as it has been shown for atmospheric channels~\cite{Andrii}. 

The theory of passive spectral filters was developed almost thirty years ago~\cite{Vogel:86,Cresser,PhysRevA.36.3803}. More recently, the topic has regained  attention due to improved experimental possibilities~\cite{Silberhorn}. New theoretical approaches have been devised, \textsl{e.g.}~\cite{Laussy}, and  different filter techniques have been compared which are usually used in optical experiments  to verify nonclassical effects of light in the spectral domain~\cite{Haeggblad}. 

For spectrally varying absorption, the transmitted fields become convolved with the response of the medium, leading to a rich structure of physical effects. A limited transmission often yields diminished field correlations. Nonclassical light is not only charcterized by correlations, but also by their variances. In general, it is not obvious, how particular quantum effects are affected by an absorptive environment.
Recently, we explained the structures of fluorescence spectra~\cite{QW-spectra} 
observed from excitons in a semiconductor quantum well structure~\cite{Burau-1,Burau-3}. 
Such systems are an interesting playground for studying effects of absorption on the quantum properties of light.

In this contribution we study the relation between the bare excitonic fluorescence fields and the actually detected fields of the quantum-well fluorescence. Second-order moments of field operators are calculated and applied to analyze squeezing. Enhancements of squeezing by absorption are demonstrated, even when the bare excitonic flourescence is no longer squeezed.

The paper is organized as follows. In section~\ref{sec.Theory} we briefly review the theory of spectral filtering of quantum light. Some of our results concerning the influence of the absorption spectra on the emission of semiconductor quantum wells are given in~section~\ref{sec.Spec}. Afterwords, in section~\ref{sec.QWcorr} we derive a general algorithm to obtain the quantum correlations of the light emitted from quantum wells. Based on the calculated moments in this section, in section~\ref{sec.sq} we study squeezing and its dependence on the absorption spectra. Finally, in section~\ref{sec.Conc} we provide a brief summary and some conclusions. 

\section{Theory of spectral filtering by absorption} \label{sec.Theory}
Due to the preservation of the canonic commutation relations, absorption acting of light induces quantum noise. Hence, in the spectral domain, the incoming field $\hat{\tilde E}_\mathrm i(\omega)$, influenced by transmission $t(\omega)$, transforms to
\begin{equation}
 \hat{\tilde E}_\mathrm f^{(+)}\hspace{-0.3cm}(\omega)=t(\omega)\hat{\tilde E}_\mathrm i^{(+)}\hspace{-0.3cm}(\omega) +\hat{\tilde E}_\mathrm n^{(+)}\hspace{-0.3cm}(\omega),\quad\hat{\tilde E}_\mathrm f^{(-)}\hspace{-0.3cm}(\omega)=\left[\hat{\tilde E}_\mathrm f^{(+)}\hspace{-0.3cm}(\omega)\right]^\dagger,\label{eq.filtertransgen}
\end{equation}
where $\hat{\tilde E}_\mathrm f$ and $\hat{\tilde E}_\mathrm n$ are the filtered and the induced noise fields, respectively.  The superscripts $(+/-)$ label positive/negative frequency parts. The tilde indicates the Fourier transform of the Heisenberg operator $\hat E(t)$. If field correlations are measured in normally and time-ordered form~\cite{WelVo}, the  induced noise plays no role, \textsl{e.g.}, in case of homodyne correlation measurements~\cite{EvgenMeas06}. Thus, we can restrict the analysis of correlations to the actual effect of the absorption spectrum on the incoming field. 

Absorption diminishes the outgoing light fields. In general, the frequency dependence of the absorption spectrum makes the calculation of these fields very difficult. For multiple frequency dependent operators it requires multitime--quantum correlations, determined via the quantum-regression theorem (QRT)~\cite{Lax}. Yet, such calculations become very simple in case of constant transmission. This is usually a good approximation, when the relevant frequency range of the influenced light field is far off-resonant to the medium.
In this case, normally-ordered field moments read as 
\begin{equation}
  \langle\hat{E}_\mathrm f^{(-)k}\hat{E}_\mathrm f^{(+)n}\rangle=t^{*k}t^n\langle\hat E_\mathrm i^{(-)k}\hat E_\mathrm i^{(+)n}\rangle,
\end{equation}
reducing the influence of the medium to a scaling.
In particular, for $k=n$ we get 
\begin{equation}
  \langle\hat{E}_\mathrm f^{(-)n}\hat{E}_\mathrm f^{(+)n}\rangle=|t|^{2n}\langle\hat E_\mathrm i^{(-)n}\hat E_\mathrm i^{(+)n}\rangle<\langle\hat E_\mathrm i^{(-)n}\hat E_\mathrm i^{(+)n}\rangle, \label{eq.constTrans}
\end{equation}
for $|t|<1$.
All functions of moments with the same number of field operators are thus scaled with the same prefactor. 

At this point, let us consider nonclassicality criteria based on moments~\cite{EvgenNC05,vo-08}. A light field is nonclassical if there exists an operator function $\hat F$ depending on field quantities,  with  $\langle\tno\hat F^\dagger\hat F\tno\rangle<0$, where $\tno\ldots\tno$ denotes time- and normal ordering. Rewriting this criterion, nonclassicality can be certified by  negativities of minors of moments, for details see~\cite{WelVo}. All summands in such a minor have the same number of positive and negative frequency field operators, resulting in the same scaling,
\begin{equation}
  \langle \tno\hat F_\mathrm f^\dagger\hat F_\mathrm f\tno\rangle=|t|^{\ell}\langle \tno\hat F_\mathrm i^\dagger\hat F_\mathrm i\tno\rangle,\quad \ell\in\mathbb{N}.
\end{equation}
This yields two conclusions. First, the range of parameters, for which nonclassicality is detected by such criteria, does not change. Second, the absolute values of the correlation functions are diminished. Hence, it should be harder to detect them. In some cases, the amplitude of the criteria can also measure the strength of  nonclassical effects. Then the nonclassical effects are also suppressed. This holds for the example of squeezing, where one measures the normally ordered field variance,
\begin{equation}
  \langle :(\Delta\hat E_\mathrm f)^2:\rangle=|t|^{2}\langle :(\Delta\hat E_\mathrm i)^2:\rangle,
\end{equation}
with $:\ldots:$ indicating normal ordering.

\section{Spectral absorption}\label{sec.Spec}
The quantum-well structure studied in~\cite{QW-spectra} will be briefly reconsidered here. A GaAs quantum well inside a multiwell stack was quasi-resonantly excited with a cw-laser of frequency $\omega_\mathrm L$. Due to the roughness of the well surface, the created excitons localize in small exciton spots (ESs)~\cite{Langbein1}, which are clearly visible after cleaning the signal from the laser background scattering. More details on the experiments can be found in Refs~\cite{Burau-1,Burau-3,Manzke}. 

Within one ES, the $N$ excitons were modeled as bosonic particles with the exciton-exciton interaction described via a nonlinear coupling~\cite{Roslyak,SpFe,Parks}. All excitons have the same resonance frequency $\omega_\mathrm x=\omega_\mathrm L+\delta$, the same spontaneous emission rate $\Gamma$, and the same coupling strength to the laser. We derived an effective Hamiltonian and a master equation~\cite{QW-spectra}, 
\begin{eqnarray}
  \hat H &=\hbar\delta\hat A^\dagger\hat A+\hbar\Omega_\mathrm R(\hat A{+}\hat A^\dagger)+\hbar G\hat A^{\dagger2}\hat A^2,\label{eq.S-Ham-2}\\
   \dot{\hat\varrho}&=\frac{1}{\rmi\hbar}[\hat H,\hat\varrho]+\frac{\Gamma}{2}(2\hat A\hat\varrho\hat A^\dagger-\hat A^\dagger\hat A\hat\varrho-\hat\varrho\hat A^\dagger\hat A).\label{eq.master-2}
\end{eqnarray}
Here we have applied collective exciton operators $\hat A$, which obey the bosonic commutation relation, $[\hat A,\hat A^\dagger]=1$. The system is described in the frame rotating with $\omega_\mathrm L$; $\Omega_\mathrm R$ and $G$ are the collective Rabi frequency and coupling strength of the exciton-exiton interaction, respectively. 

The Hamiltonian in equation~(\ref{eq.S-Ham-2}) describes the dynamics of the excitons which have been excited by the driving laser field. 
However, it does not describe the light fields detected outside of the quantum well. In order to emit photons the medium must first absorb the laser photons, which means, that higher absorption yields higher exciton densities and in turn higher emission intensities. Hence, the quantum-well emission spectrum depends both on the emission spectrum of the excitons and the absorption spectrum. This result~\cite{QW-spectra} is consistent with both the input-output formalism of quantum optics~\cite{Dima1} and Kirchhoffs law of radiation in non-equilibrium many-body systems~\cite{Henneb1}. Thus,  in our system absorption dominates the  transmission properties.

From the above discussion it follows, that we are in fact dealing with two kinds of fields. On one hand, we have the field from the bare excitons, as derived by solving equation~(\ref{eq.master-2}). It represents the incoming field transmitting the semiconductor. It should be clearly stated that, in the case of a semiconductor quantum well, these fields do not have a physical reality. The measured fields are emitted from the quantum well and include the influence of the absorption of the medium. To distinguish the different fields and their correlations, we will denote them by exciton fluorescence (index 'x') and quantum-well fluorescence (index 'q'), respectively.

The exciton-emission spectrum $S_\mathrm x(\omega)$ can be calculated using the Wiener-Khintchine theorem~\cite{Wiener} with the source field given by the collective exciton operators, and the QRT~\cite{WelVo}. The absorption spectrum $a(\omega)$ follows from the input-output formalism~\cite{Dima1}, which relates it to the susceptibility $\chi(\omega)$. We apply a simple oscillator model,
\begin{equation}
  \chi(\omega)=\frac{f}{\omega-\omega_\mathrm x-\rmi\frac{\Gamma}{2}}=\frac{f}{\omega-\omega_\mathrm L-\delta-\rmi\frac{\Gamma}{2}},\label{eq.susz}
\end{equation}
where $f$ is the oscillator strength. The susceptibility relates the transmission and reflection coefficient, $t(\omega)$ and $r(\omega)$, respectively, to the absorption spectrum,
\begin{equation}
   a(\omega)=1-|t(\omega)|^2-|r(\omega)|^2.\label{eq.absorp}
\end{equation}
Finally, the quantum-well spectrum $S_\mathrm q(\omega)$ reads as
\begin{equation}
   S_\mathrm{q}(\omega)=a(\omega)S_\mathrm x(\omega).\label{eq.QWspec}
\end{equation}

\begin{figure}[h]
    \includegraphics[width=8cm]{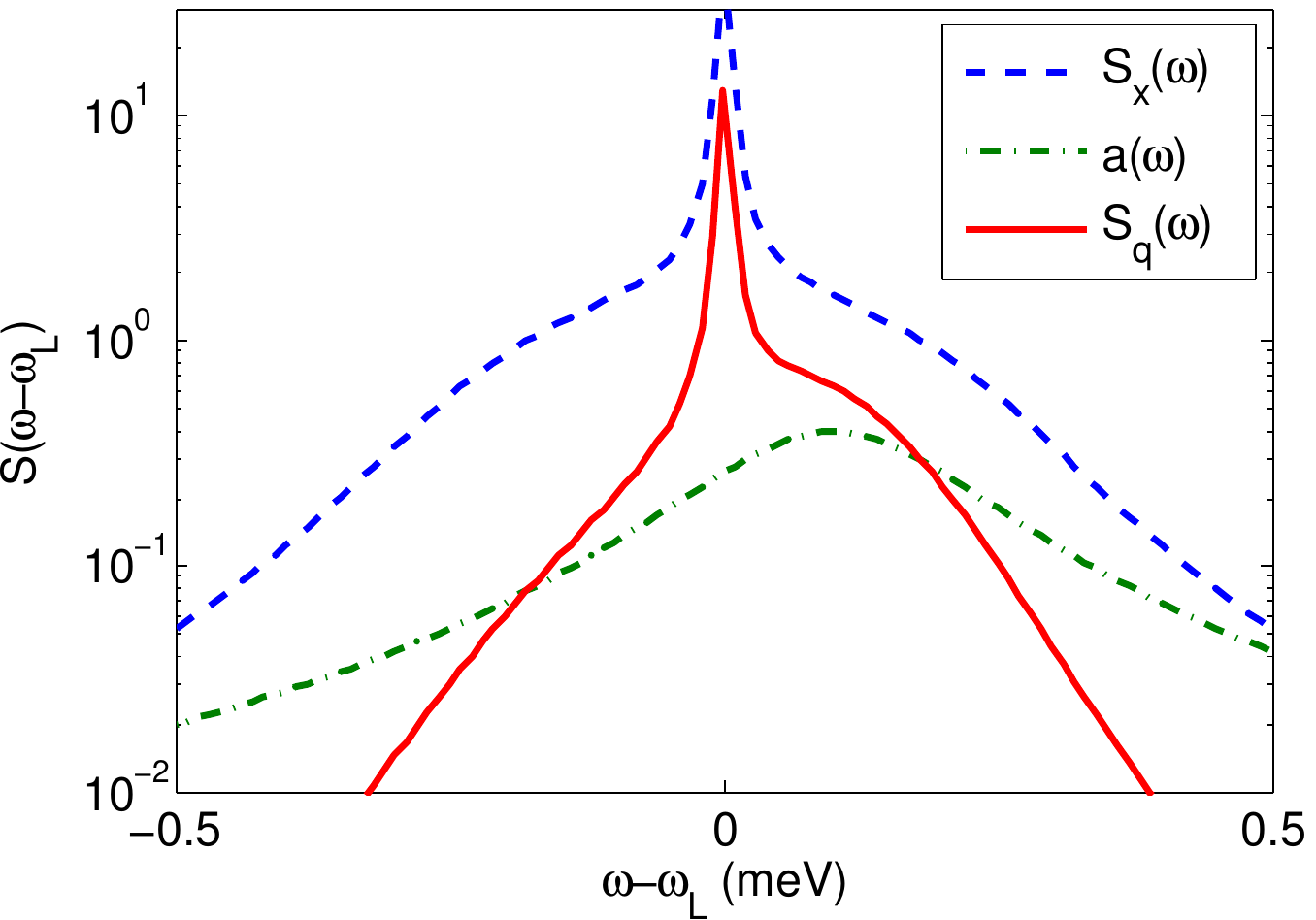}
    \caption{The different contributions to the quantum-well fluorescence spectrum: $S_\mathrm x(\omega)$ (blue, dashed), $a(\omega)$ (green, dot-dashed), and $S_\mathrm q(\omega)$ (red, solid). The parameters are: $G=0.15$ meV, $\Gamma=0.2$ meV, $\Omega_\mathrm R=0.1$ meV, $\delta=0.1$ meV, $f=1$ meV. The spectral resolution of the detector is modeled with a width $\Gamma_\mathrm f=0.0107$ meV.}\label{fig.QWfluor}
\end{figure}

All three spectra are shown in figure~\ref{fig.QWfluor} for the parameters given in the caption. The combination of exciton dynamics and medium response leads to the asymmetric shape of the measured spectra. For a more detailed interpretation of the spectra see~\cite{QW-spectra}. The set of fit parameters obtained from the simulations is listed in table~\ref{tab.fit}. 
We note that such spectra have been recorded using GaAs quantum wells cooled down to $T=4.5$~K, cf.~\cite{QW-spectra}. The following study of quantum effects is based on the same model, as the disturbances at higher temperatures are expected to diminish the quantum signatures of the emitted light.

\begin{table}[h]
   \begin{tabularx}{0.45\textwidth}{XXXl}\hline\hline
    $P_{\mathrm L}$ ($\mu$W) & 100 & 150 & 310\\
      \hline
    $\hbar G$ (meV) & 0.10& 0.205& 0.45\\
    $\hbar\Omega_{\mathrm R}$ (meV) & 0.045& 0.075& 0.16\\
    $\hbar\delta$ (meV) & 0.08& 0.08& 0.09\\
    $f$ (a.u.) & 1.0& 1.0& 0.9\\
    $\hbar\Gamma$ (meV) & 0.15& 0.20& 0.22\\
      \hline\hline
  \end{tabularx}
\caption{Fit parameters of the spectra measured for different laser powers, after~\cite{QW-spectra}. 
}\label{tab.fit}
\end{table}

\section{Quantum Correlation Functions}\label{sec.QWcorr}
\subsection{General Correlation functions}
In the following, we deduce general quantum-well correlations from the excitonic fluorescence correlations and the absorption. We want to apply the input-output formalism developed in~\cite{Gruner,Dima1}. The source fields in time and frequency domain are related as
\begin{eqnarray}
   \hat E_\mathrm{x,s}^{(+)}\propto\hat A,\quad  
   \hat{\tilde E}^{(+)}_\mathrm{x,s}(\omega)\propto\hat b(\omega),\label{eq.excSource}
\end{eqnarray}
where $\hat b(\omega)$ are mode densities in the continuum of frequencies $\omega$.
Then the results for the spectrum indicate in particular that 
\begin{equation}
   I_\mathrm q=\langle\hat A^\dagger\hat A\rangle_\mathrm q=\int\limits_{-\infty}^\infty \rmd\omega\,S_\mathrm q(\omega)=\int\limits_{-\infty}^\infty \rmd\omega\,a(\omega)S_\mathrm x(\omega).\label{eq.Intexcout}
\end{equation}
and more generally, that
\begin{eqnarray}
    \hat{\tilde E}^{(+)}_\mathrm{q,s}(\omega)&\propto\sqrt{a(\omega)}\hat b(\omega),\quad \hat{\tilde E}^{(-)}_\mathrm{q,s}(\omega)=\left[\hat{\tilde E}^{(+)}_\mathrm{q,s}(\omega)\right]^\dagger.\label{eq.relExcQW}
\end{eqnarray}
The quantum-well fluorescence fields are thus described by convolutions between the Heisenberg operators $\hat A(t)$ and the Fourier transform of $\sqrt{a(\omega)}$. In the full field operators one has to add the free field parts. However, as we only consider time- and normal-ordered correlation functions and assume that only the source field hits the detector, the free fields do not contribute to the quantum expectation values~\cite{WelVo}. Hence we omit the index 's' in the following.

The algorithm to obtain the quantum-well fluorescence correlations from the corresponding exciton fluorescence correlations is the same as for spectrally filtered fields. For simplicity we limit the discussion to steady-state correlations. Let $f_\mathrm x(0)$ be a general steady-state exciton correlation function,
\begin{equation}
   f_\mathrm x(0)=\langle\tno\hat A^{\dagger m}\hat A^n\tno\rangle,\quad m,n\in\mathbb N,
\end{equation}
calculated by solving the master equation~(\ref{eq.master-2}). The corresponding quantum-well fluorescence correlation function $f_\mathrm q(0)$ is an $m+n$-dimensional convolution with the Fourier-transform of the square root of the absorption spectrum, $\sqrt{a(\omega)}$. The time-dependent function $f_\mathrm x(\{t_j\})$ is given by transforming each field operator in $f_\mathrm x(0)$ into a Heisenberg operator of different time $t_j$,
\begin{equation}
   f_\mathrm x(\{t_j\})=\langle\tno\prod\limits_{j=1}^m\hat A^{\dagger}(t_j)\prod\limits_{j=m+1}^{m+n}\hat A(t_j)\tno\rangle,
\end{equation}
using the QRT. A multi-dimensional Fourier-transform  is performed, with different frequencies $\omega_j$ used for each operator, yielding
\begin{eqnarray}
	\tilde f_\mathrm x(\{\omega_j\})=&\int\limits_{-\infty}^\infty& \rmd t_1\ldots \rmd t_{m+n}\times\nonumber\\
	&&\rme^{\rmi[\sum\limits_{j=1}^m\omega_jt_j-\hspace{-0.1cm}\sum\limits_{j=m+1}^{m+n}\omega_jt_j]}f_\mathrm x(\{t_j\}).
\end{eqnarray}
The Fourier-transformed function $\tilde f_\mathrm x(\{\omega_j\})$ is scaled as 
\begin{equation}
	\tilde f_\mathrm q(\{\omega_j\})=\sqrt{a(\omega_1)\ldots a(\omega_{m+n})}\tilde f_\mathrm x(\{\omega_j\}),
\end{equation}
which is the spectral density of the quantum-well fluorescence correlation. Finally, this correlation is integrated over all frequencies to obtain 
\begin{equation}
  f_\mathrm q(0)=\frac{1}{(2\pi)^{m+n}}\int\limits_{-\infty}^\infty \rmd\omega_1\ldots \rmd\omega_{m+n}\,\tilde f_\mathrm q(\{\omega_j\}).\label{eq.genTransf}   
\end{equation}

We point out, that no specifics were included about the Hamiltonian, or the absorption. Therefore, this formalism can be applied also to more general situtations. 
Furthermore, we note, that for frequency independent absorption, equation~(\ref{eq.genTransf}) reduces to the known scaled version of the correlation functions
\begin{equation}
  f_\mathrm q(0)=a^\frac{m+n}{2}f_\mathrm x(0).   
\end{equation}

\subsection{Moments of second order}
Due to the limited number of available spectra studied in~\cite{QW-spectra}, we have only the three data sets in table~\ref{tab.fit}. In order to better illustrate the evolution of the correlations with increasing laser power, we interpolate the system quantities between the measurements 1 and 3, using natural cubic splines. All expectation values and correlation functions are depicted between the experimentally studied pump-laser powers from $P_\mathrm L=100$ $\mu$W to 310 $\mu$W.

The simplest expectation value is the coherent part of the intensity of the quantum-well emission, which is proportional to $|\langle\hat A\rangle_\mathrm q|^2$, the latter will be simply denoted as coherence. In the steady-state regime we obtain 
\begin{eqnarray}
   \langle\hat A\rangle_\mathrm{q}&=\frac{1}{2\pi}\int \rmd\omega\,\sqrt{a(\omega)}\int \rmd t\,\rme^{-\rmi\omega t}\langle\hat A(0)\rangle_\mathrm x\nonumber\\
   &=\sqrt{a(0)}\langle\hat A(0)\rangle_\mathrm x.
\end{eqnarray}
Here and in the following, integrals from $-\infty$ to $\infty$ will be written without borders. Due to the steady state situation, the $t$-integration yields a $\delta$-function. The frequency zero in the absorption represents $\omega_\mathrm L$. The coherent part of the intensity then follows as
\begin{equation}
   |\langle\hat A\rangle_\mathrm{q}|^2=a(0)|\langle\hat A\rangle_\mathrm x|^2.\label{eq:coh-int}
\end{equation}
The coherence is visible in the Rayleigh-peak at the laser frequency, and hence, it only scales with the absorption at $\omega_\mathrm L$, as given in equation~(\ref{eq:coh-int}). 

The intensity of the quantum-well fluorescence, $I_\mathrm q \propto \langle\hat A^\dagger\hat A\rangle_\mathrm q$, was already discussed above. 
With the present algorithm we get
\begin{eqnarray}
   I_\mathrm x(\{t_j\})&=\langle\hat A^\dagger(t_1)\hat A(t_2)\rangle_\mathrm x,\\
    \tilde I_\mathrm x(\{\omega_j\})&=\int \hspace{-0.1cm} \rmd t_1 \hspace{-0.1cm}\int  \hspace{-0.1cm}\rmd t_2\,\rme^{\rmi(\omega_1t_1-\omega_2t_2)}\langle \hat A^\dagger(t_1)\hat A(t_2)\rangle_\mathrm x.
\end{eqnarray}
Setting $\tau=t_2-t_1\mathrm{,\ }\Delta=\omega_1-\omega_2$ yields
\begin{eqnarray}
   I_\mathrm x(\{t_j\}) &=2\pi \delta(\Delta)\int \rmd\tau\,\rme^{-\rmi\omega_2\tau}\langle \hat A^\dagger(0)\hat A(\tau)\rangle_\mathrm x\label{eq.steady}\nonumber\\
    &=(2\pi)^2\delta(\Delta)S_\mathrm{x}(\omega_2),\\
   I_\mathrm{q}(0)&=\frac{1}{(2\pi)^2}\int \rmd\omega_1 \int \rmd\omega_2 \,\tilde I_\mathrm q(\{\omega_j\})\nonumber\\
     &=\int \rmd\omega_1\,a(\omega_1)S_\mathrm x(\omega_1).
\end{eqnarray}
We used the fact that the system is in the steady state in equation~(\ref{eq.steady}), thus reobtaining eq.~(\ref{eq.Intexcout}).

\begin{figure}[h]
\includegraphics[width=7cm]{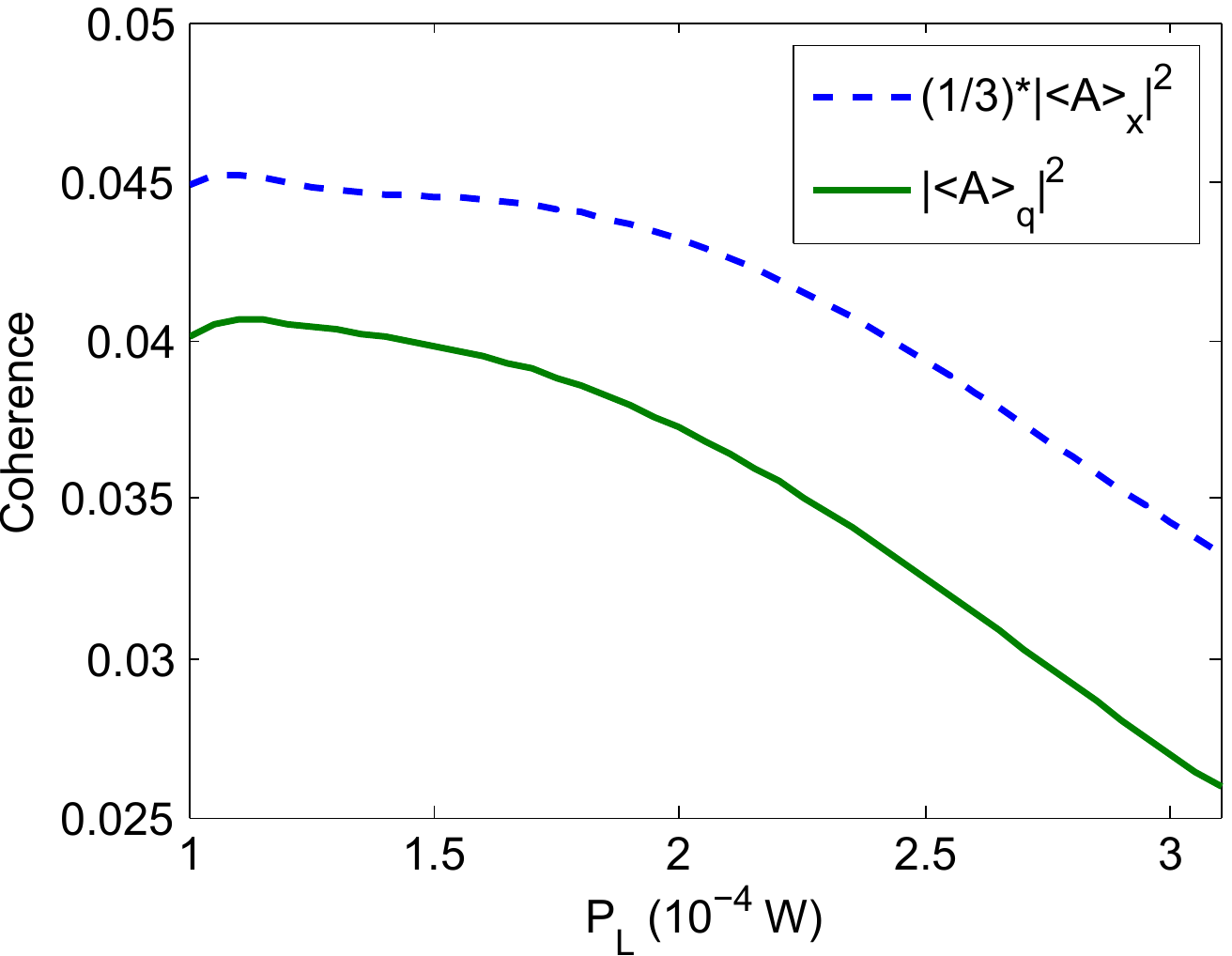}
\includegraphics[width=7cm]{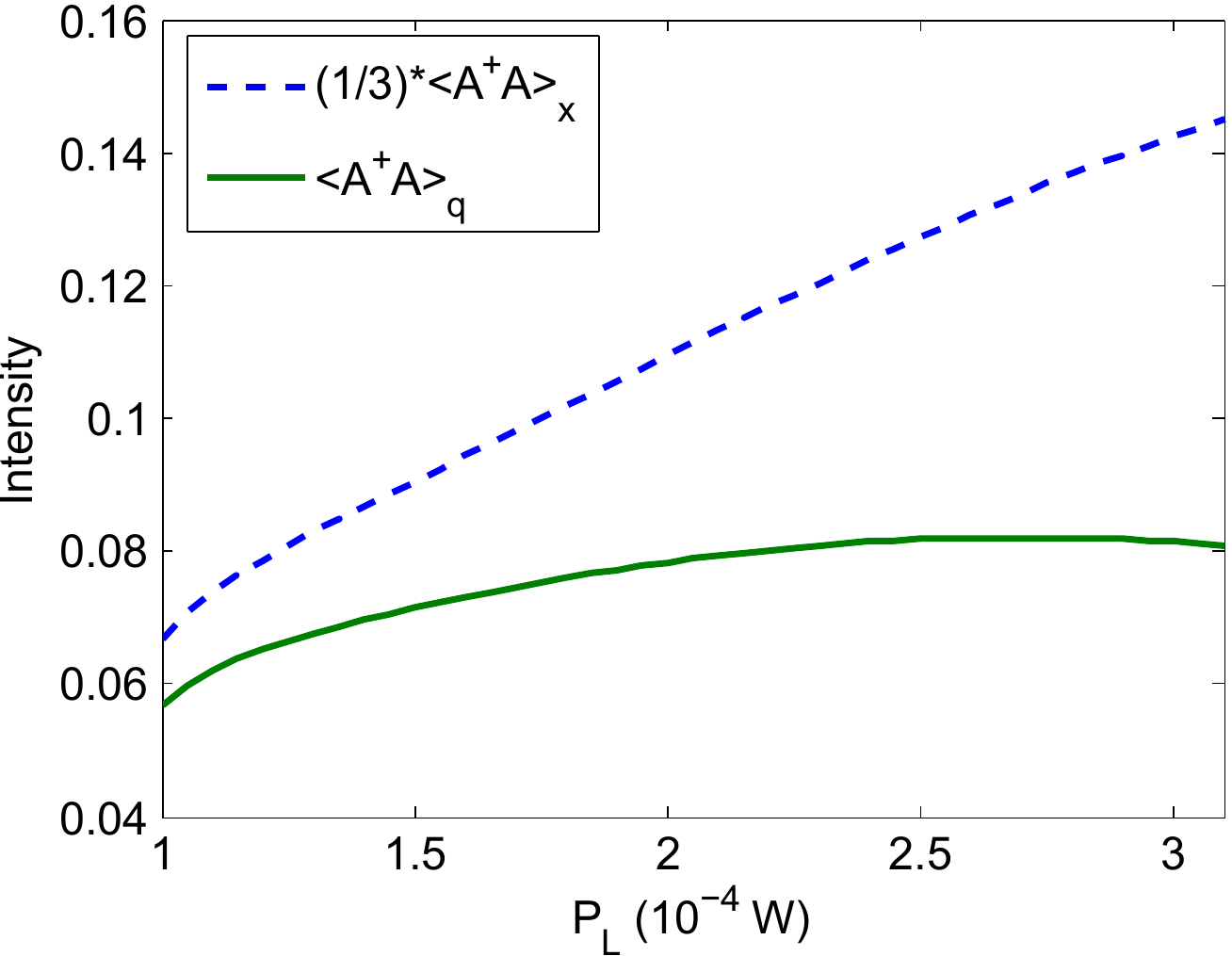}
\caption{(Color online) coherence (top) and intensity (bottom) of the exciton (dashed) and quantum-well emission (solid) for increasing $P_\mathrm L$.}\label{fig.int-coh-in-out}
\end{figure}

In figure~\ref{fig.int-coh-in-out} we compare the intensity of the emission and its coherent part. 
Both, the exciton- and the quantum-well-fluorescence coherence decrease with increasing laser power. The exciton-fluorescence intensity increases almost linearly with increasing laser power. Hence, the emission for higher excitation becomes more and more incoherent. The quantum-well-fluorescence intensity, however, increases much slower and even approaches a maximum at some point. We conclude, that the degree of coherence, defined as
\begin{equation}
   D_\mathrm{coh}=\frac{|\langle\hat A\rangle|^2}{\langle\hat A^\dagger\hat A\rangle},
\end{equation}
for the quantum-well fluorescence is actually larger than the corresponding value of the exciton fluorescence. Hence, while the coherence is diminshed by absorption, the corresponding degree of coherence may increase.

Let us analyze the anomalous moment $\langle\hat A^2\rangle$ of second order. In this case, the normal ordering and the positive time argument in the QRT become relevant. With $f_\mathrm x(0)=\langle\hat A^2\rangle_\mathrm x$, we find for the anomalous moment of the quantum-well fluorescence
\begin{eqnarray}
	f_\mathrm x(\{t_j\})&=&\langle\tno\hat A(t_1)\hat A(t_2)\tno\rangle_\mathrm x,\\
	\tilde f_\mathrm x(\{\omega_j\})&=&\!\int \rmd t\, \rme^{\rmi(\omega_1+\omega_2)t}\int \rmd\tau\,\rme^{\rmi\omega_2\tau}\langle\tno\hat A(\tau)\hat A(0)\tno\rangle_\mathrm x\nonumber\\
	&=&\!4\pi\delta(\omega_1+\omega_2)\hspace{-0.1cm}\int\limits_{0}^\infty\hspace{-0.1cm} \rmd\tau\cos(\omega_2\tau)\langle\hat A(\tau)\hat A(0)\rangle_\mathrm x,\\
	\langle\hat A^2\rangle_\mathrm{q}&=&\frac{1}{\pi}\int \rmd\omega_2\,\sqrt{a(\omega_2)a(-\omega_2)}\times\nonumber\\
	&&\int\limits_{0}^\infty \rmd\tau\cos(\omega_2\tau)\langle\hat A(\tau)\hat A(0)\rangle_\mathrm x.
\end{eqnarray}
The absorption frequencies are now correlated symmetrically around $\omega_\mathrm L$.

\begin{figure}[h]
\includegraphics[width=7cm]{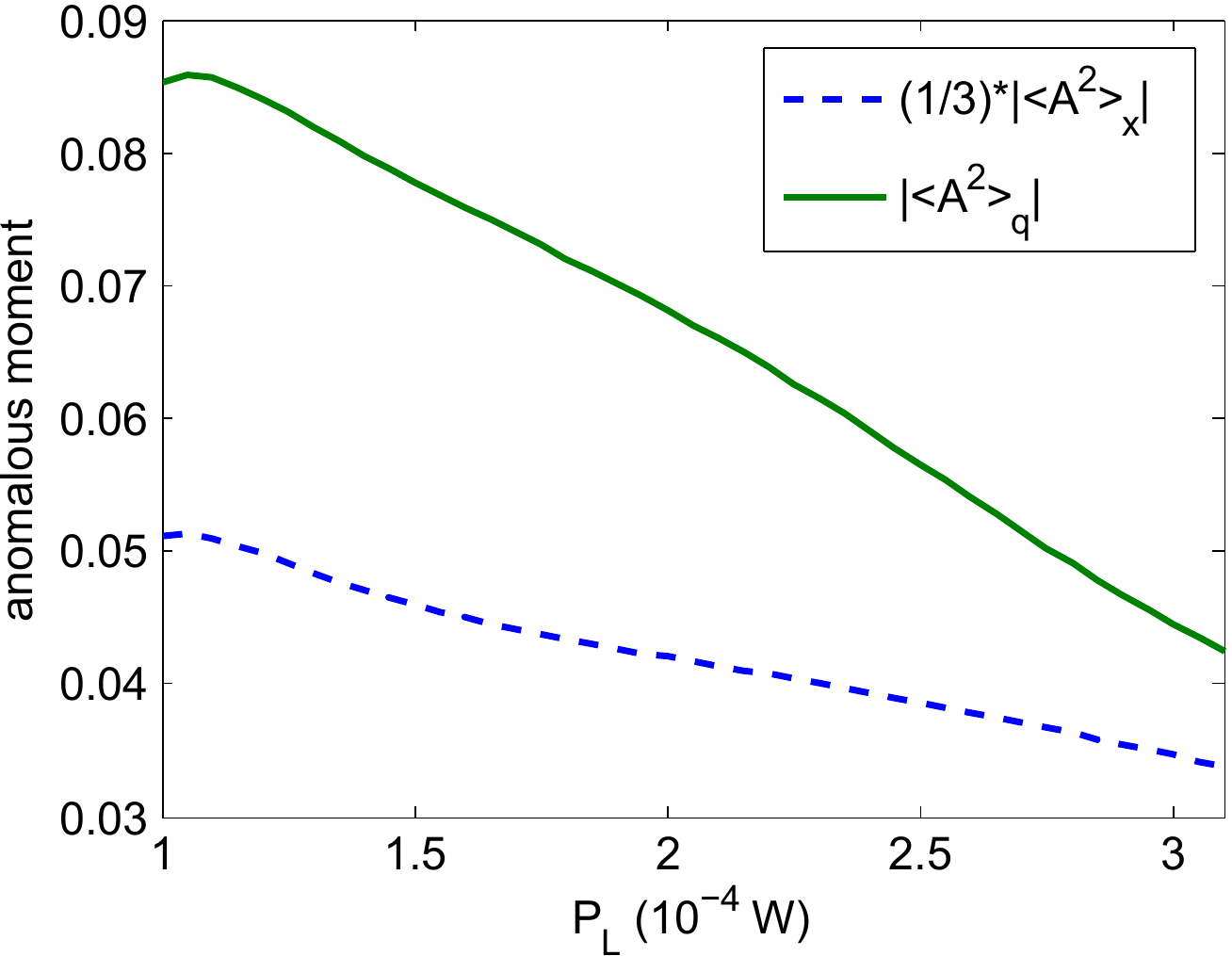}
\caption{(Color online) Modulus of the anomalous moment of exciton (dashed) and quantum-well emission (solid) for increasing $P_L$.}\label{fig.A2-in-out}
\end{figure}

The modulus of the anomalous second moment, $|\langle\hat A^2\rangle|$, is depicted in figure~\ref{fig.A2-in-out}. Absorption yields a stronger decay of the quantum-well moment compared to the excitonic one. However, this suppression is much weaker than those of coherence and intensity.

\section{Squeezing}\label{sec.sq}
Based on the intensity, the coherence and the anomalous moment as given above, we are able to study squeezing in the exciton- and quantum-well fluorescence. A light field is squeezed, if the normally-ordered field variance becomes negative, corresponding to field fluctuations below the vacuum noise level~\cite{WelVo}. Scaling the coupling between the source field and the operator $\hat A$ with $\zeta$, the phase-optimized, normally-ordered field variance of the exciton fluorescence becomes
\begin{equation}
	\frac{\langle:(\Delta\hat E_\mathrm x)^2:\rangle}{|\zeta|^2}=2(\langle\hat A^\dagger\hat A\rangle_\mathrm x{-}|\langle\hat A\rangle_\mathrm x|^2{-}|\langle \hat A\rangle_\mathrm x^2-\langle \hat A^2\rangle_\mathrm x|).\label{eq.sq-exc}
\end{equation}
The corresponding normally-ordered field variance of the quantum-well fluorescence reads as
\begin{equation}
	\frac{\langle:(\Delta\hat E_\mathrm q)^2:\rangle}{|\zeta|^2}=2(\langle\hat A^\dagger\hat A\rangle_\mathrm q{-}|\langle\hat A\rangle_\mathrm q|^2{-}|\langle \hat A\rangle_\mathrm q^2-\langle \hat A^2\rangle_\mathrm q|).\label{eq.sq-well}
\end{equation}

\begin{figure}[h]
\centering
\includegraphics[width=7cm]{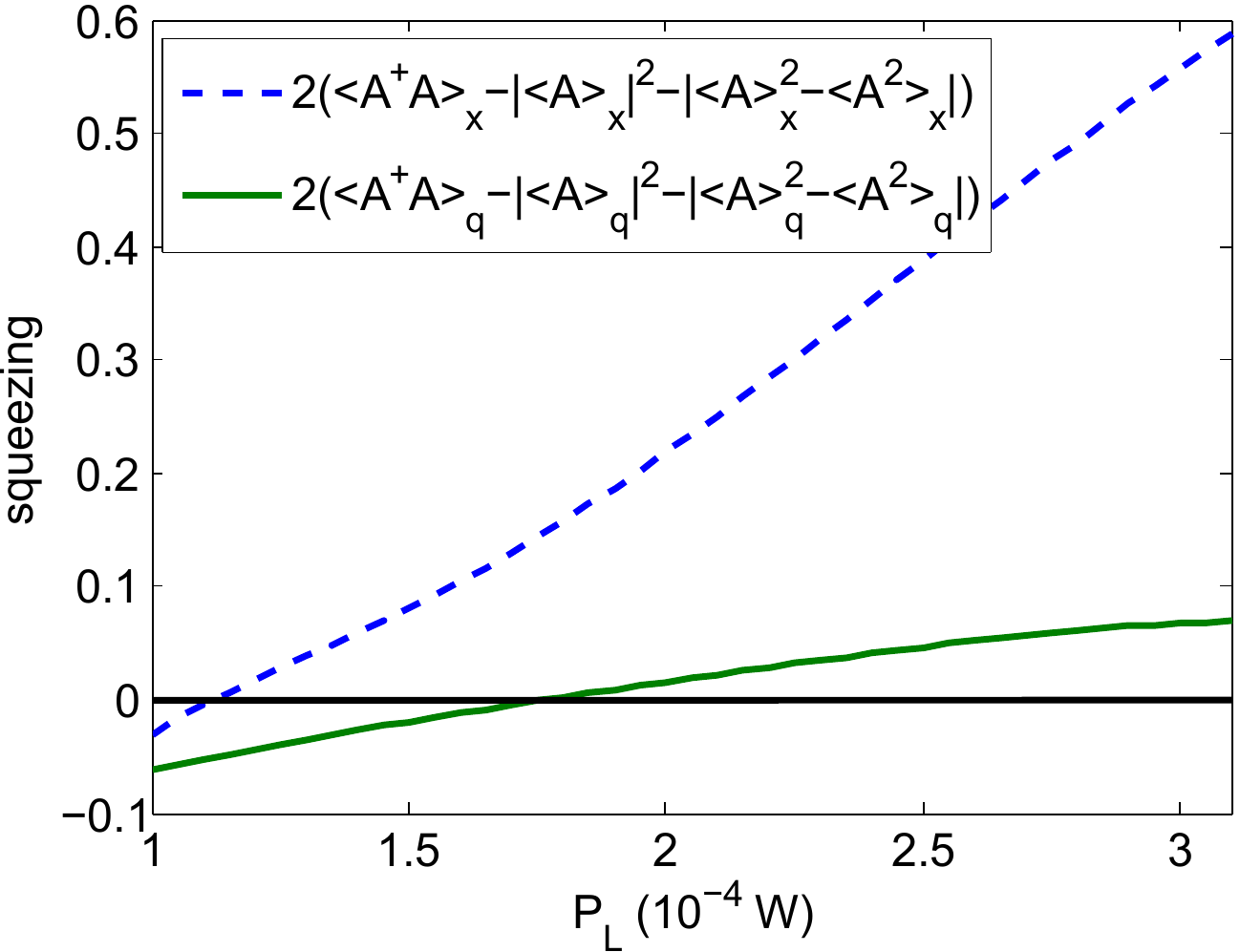}
\caption{(Color online) Normally-ordered field variance of the exciton (dashed) and the corresponding quantum-well emission (solid)  for increasing $P_\mathrm L$.}\label{fig.sq-in-out}
\end{figure}

In figure~\ref{fig.sq-in-out} we compare the squeezing of the exciton- and the quantum-well-fluorescence. Squeezing of the exciton fluorescence is limited to very low laser powers, as for higher exciton densities the incoherence of the emission becomes dominant. In the quantum-well fluorescence, on the other hand, the squeezing persists up to a higher laser power. For all considered laser powers, we find
\begin{equation}
	\langle:(\Delta\hat E_\mathrm q)^2:\rangle<\langle:(\Delta\hat E_\mathrm x)^2:\rangle,
\end{equation}  
so that the fluctuations of the light emitted by the quantum well are smaller than those of the bare excitonic emission.
This may seem surprising, as the latter field is convolved with the absorption spectrum to derive the quantum-well field. However, the coherence is only affected by the absorption at the laser frequency, while the intensity is a convolution with the full absorption spectrum. Both, the coherence and the anomalous moment are less suppressed by the absorption than the intensity. Consequently, due to the spectral absorption properties, the quantum-well fluorescence shows stronger squeezing than the bare excitonic emission. 
We thus expect that an off-resonant filter, with a filter function corresponding to $\sqrt{a(\omega)}$, may yield enhanced squeezing also in other experimental scenarios.

\begin{figure}[h]
\includegraphics[width=7cm]{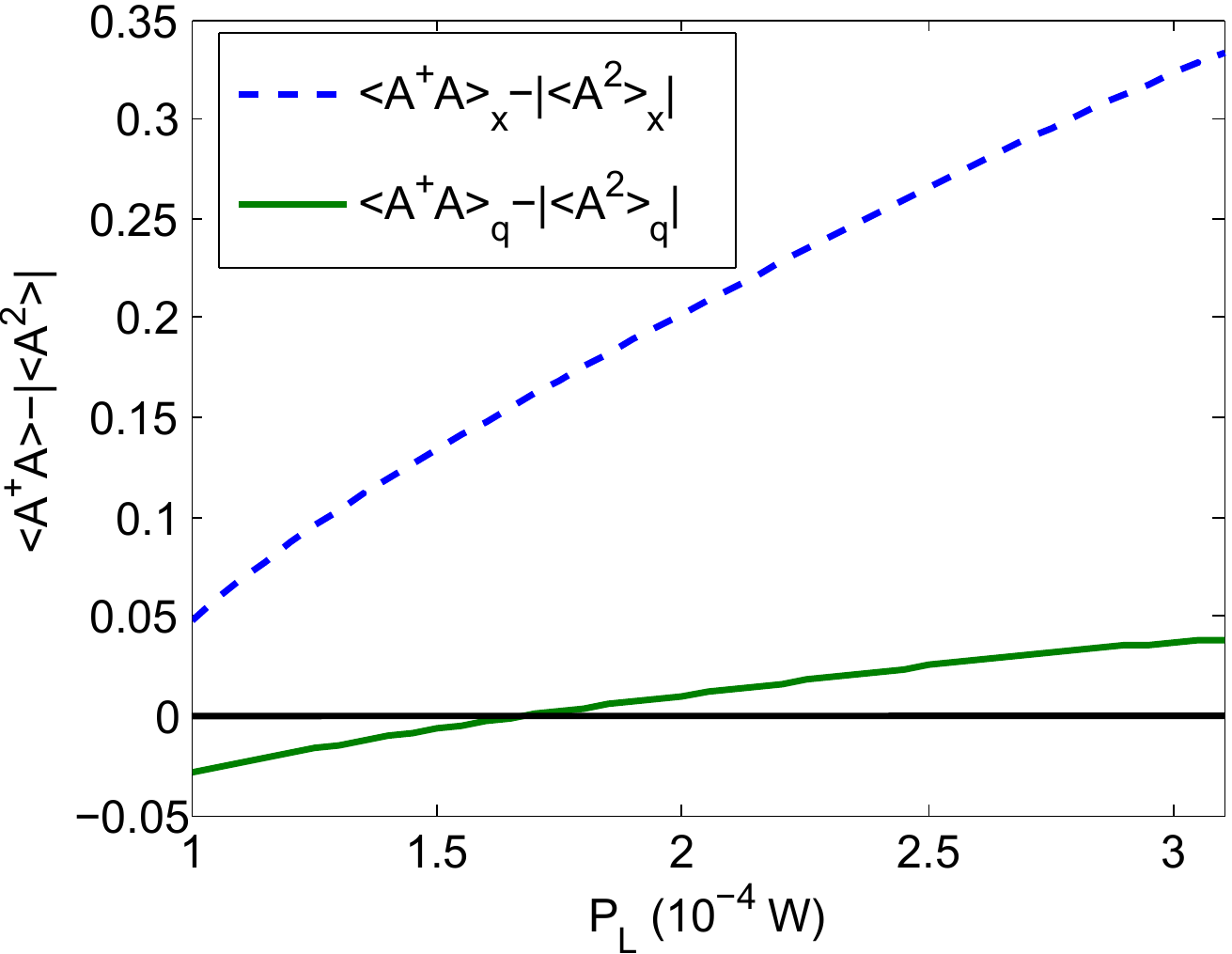}
\caption{(Color online) $\langle\hat A^\dagger\hat A\rangle-|\langle\hat A^2\rangle|$ for exciton (dashed) and quantum-well emission (solid) for increasing $P_L$.}\label{fig.QW-Int-A2}
\end{figure}

For a light source with a vanishing coherent amplitude, $\langle\hat A\rangle=0$, the squeezing condition simplifies to
\begin{equation}
   \langle\hat A^\dagger\hat A\rangle-|\langle\hat A^2\rangle|<0.\label{eq.A2NC}
\end{equation}
In our case $\langle \hat A\rangle$ is not zero and equation~(\ref{eq.A2NC}) represents a different nonclassicality condition, cf.~\cite{EvgenNC05}.
As the second-order anomalous moment of the quantum-well fluorescence is less suppressed by absorption than the intensity, this condition is interesting on its own. Even when the exciton fluorescence does not show this nonclassicality, it occurs in the quantum-well fluorescence for sufficiently low laser powers, see figure~\ref{fig.QW-Int-A2}. 

The two nonclassicality conditions in figures.~\ref{fig.sq-in-out},\ref{fig.QW-Int-A2} behave rather similarly for the quantum-well fluorescence. As the modulus of $\langle\hat A\rangle_\mathrm q$ is not small compared to the other contributions, this similarity indicates a phase matching between $\langle\hat A\rangle_\mathrm q^2$ and $\langle\hat A^2\rangle_\mathrm q$ as the origin of the enhanced nonclassicality. As the coherence only scales with the positive absorption $\sqrt{a(0)}$, the coherent amplitudes $\langle\hat A\rangle_\mathrm x$ and $\langle\hat A\rangle_\mathrm q$ have the same phase.

\begin{figure}[h]
   \includegraphics[width=7cm]{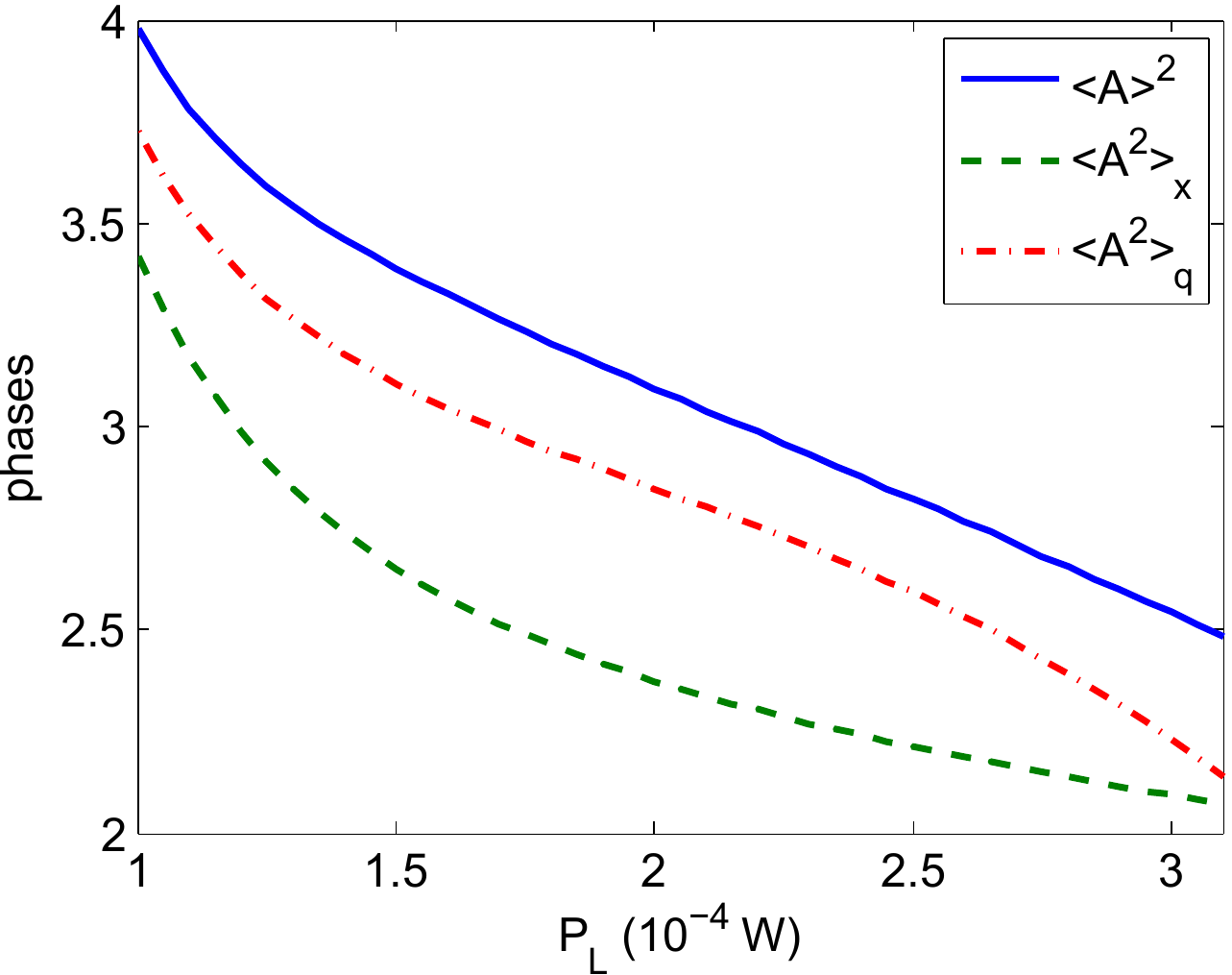}
   \caption{(Color online) Phases of $\langle\hat A\rangle^2$ (solid), $\langle\hat A^2\rangle_\mathrm x$ (dashed), and $\langle\hat A^2\rangle_\mathrm q$ (dash-dotted) for increasing $P_L$.}\label{fig.phase} 
\end{figure}

In figure~\ref{fig.phase} the phases of $\langle\hat A\rangle^2$, $\langle\hat A^2\rangle_\mathrm x$, and $\langle\hat A^2\rangle_\mathrm q$ are shown as a function of the pump power. Indeed, the phase of the second moment of the quantum-well emission, $\langle\hat A^2\rangle_\mathrm q$, is closer to the phase of $\langle\hat A\rangle^2$ than the phase of $\langle\hat A^2\rangle_\mathrm x$. Therefore, the absorption in our scenario yields an increase in phase matching, thus producing nonclassical light as defined by equation~(\ref{eq.A2NC}), and increasing the squeezing, equation~(\ref{eq.sq-well}).

\section{Summary and Conclusions}\label{sec.Conc}
We have studied the fluorescence of a GaAs quantum well. The source fields of the excitons are scaled with the absorption spectrum, as the latter is relevant for the exciton generation. The actual quantum-well fields are thus convolutions of the corresponding exciton operators and the absorption spectra. The intensity and its coherent part are suppressed by absorption. However, the absorption at the laser frequency diminishes the coherent part less than the full intensity, as the latter is affected by the full absorption spectrum. Consequently, squeezing of the quantum-well fluorescence is stronger and more persistent for higher pump powers, than the squeezing of the bare excitonic fluorescence. In addition to squeezing, we  
find another nonclassical effect, which is directly caused by a dominant anomalous moment of second order. Both effects can be explained by an increased phase matching due to the absorption.

\paragraph{Acknowledgements.}
This work was supported by Deutsche Forschungsgemeinschaft through SFB 652.

\end{document}